\documentclass[10pt,preprint]{aastex} 
\usepackage{float} 
 
\def\Ho  {\ifmmode {H_0} \else {$H_0$} \fi} 

\def\HHO  {\ifmmode {H_2O} \else {$H_2O$} \fi}
 
\def\kms {km~s$^{-1}$} 
\def\kmsmpc {km~s$^{-1}$~Mpc$^{-1}$} 
\def\kmsyr {km~s$^{-1}$~yr$^{-1}$} 
 
\begin{document}

\title{The Megamaser Cosmology Project. VI. Observations of NGC 6323 }  
 
\author{C. Y. Kuo\altaffilmark{1}, J. A. Braatz\altaffilmark{2},
K. Y. Lo\altaffilmark{2}, M. J. Reid\altaffilmark{3} , S. H. Suyu\altaffilmark{1}, 
D. W. Pesce\altaffilmark{4}, J. J. Condon\altaffilmark{2}, C. Henkel\altaffilmark{5,6},
C. M. V. Impellizzeri\altaffilmark{2} }                                 
 
\affil{\altaffilmark{1}Academia Sinica Institute of Astronomy and
Astrophysics, P.O. Box 23-141, Taipei 10617, Taiwan }                   
\affil{\altaffilmark{2}National Radio Astronomy Observatory, 520 
Edgemont Road, Charlottesville, VA 22903, USA} 
\affil{\altaffilmark{3}Harvard-Smithsonian Center for Astrophysics, 
60 Garden Street, Cambridge, MA 02138, USA} 
\affil{\altaffilmark{4}Department of Astronomy, University of Virginia, 
Charlottesville, VA 22904} 
\affil{\altaffilmark{5}Max-Planck-Institut f\"ur Radioastronomie, 
Auf dem H\"ugel 69, 53121 Bonn, Germany} 
\affil{\altaffilmark{6}Astronomy Department, Faculty of Science,
King Abdulaziz University, P.O. Box 80203,                              
Jeddah, Saudi Arabia}

\begin{abstract} 
We present observations of the H$_{2}$O megamasers in the
accretion disk of NGC 6323.  By combining interferometric and 
spectral monitoring data, we estimate $\Ho = 73^{+26}_{-22}$ km/s/Mpc, 
where the low strength of the systemic masers ($<15$ mJy) limits the 
accuracy of this estimate.  The methods developed here for dealing with 
weak maser emission provide guidance for observations of similar sources, 
until significant increases in radio telescope sensitivity, such as
anticipated from the next generation Very Large Array, are realized.
\end{abstract} 
 
 
\section{INTRODUCTION} 
H$_{2}$O megamasers provide a direct determination of $H_{0}$ independent
of standard candles (e.g. Reid et al. 2009, Braatz et al. 2010, Reid
et al. 2013, Kuo et al. 2013). This method involves sub-milliarcsecond
resolution imaging of H$_{2}$O maser emission from sub-parsec circumnuclear
disks at the center of active galaxies, and the \emph{geometric}
distance to each of the galaxies is determined based on measurements
of the orbital size and velocity as well as the centripetal acceleration
of maser clouds orbiting around the supermassive black hole. Since
this technique involves very few assumptions and can be applied to
maser galaxies in the Hubble flow in a single step, systematic
errors are expected to be small. In the Megamaser Cosmology Project
(MCP), we have currently obtained a $H_{0}$ of $68\pm7$ \kmsmpc\ 
from UGC 3789 (Reid et al. 2013) and $68\pm11$ \kmsmpc\ 
from NGC 6264 (Kuo et al. 2013). To further constrain $H_{0}$, we
are currently measuring additional maser galaxies and searching for
more high quality megamaser disk systems which are suitable for $H_{0}$
determination.                                                          
 
In this paper, we present observations of NGC 6323,
a Seyfert 2 galaxy at a distance of $\sim100$ Mpc. The masers in this galaxy shows
a Keplerian rotation curve and have the necessary maser
components (i.e. the systemic and high-velocity masers; see Kuo et
al. 2011 for their definitions) for a $H_{0}$ determination with
the H$_{2}$O megamaser technique. It is different from our previously
published maser galaxies (i.e. UGC 3789 and NGC 6264) in that the systemic 
masers, are quite weak. The typical signal-to-noise ratio (SNR) of a systemic maser 
line is only $\sim$10 in spectra taken by the Green Bank Telescope (GBT)
\footnote{The Green Bank Telescope is a facility of the National Radio Astronomy Observatory.}
with an integration time of $\approx3$ hours. Since the SNR of the systemic 
masers plays a crucial role in the precision of the $H_{0}$ estimate,
NGC 6323 provides a test case to explore the accuracy and precision
one can achieve in determining $H_{0}$ from a galaxy with faint systemic
masers. 

In section 2, we present our VLBI and single-dish observations.
In section 3, we show the analysis of the centripetal accelerations
of the masers in NGC 6323. The analyses of the Hubble constant determination
are presented in section 4. Finally, we summarize the results in
section 5.                                                              
 
\section{Observations and Data Reduction} 
 
\subsection{GBT monitoring} 

\begin{figure}[!htb] 
\begin{center} 
\includegraphics[angle=0, scale=0.6]{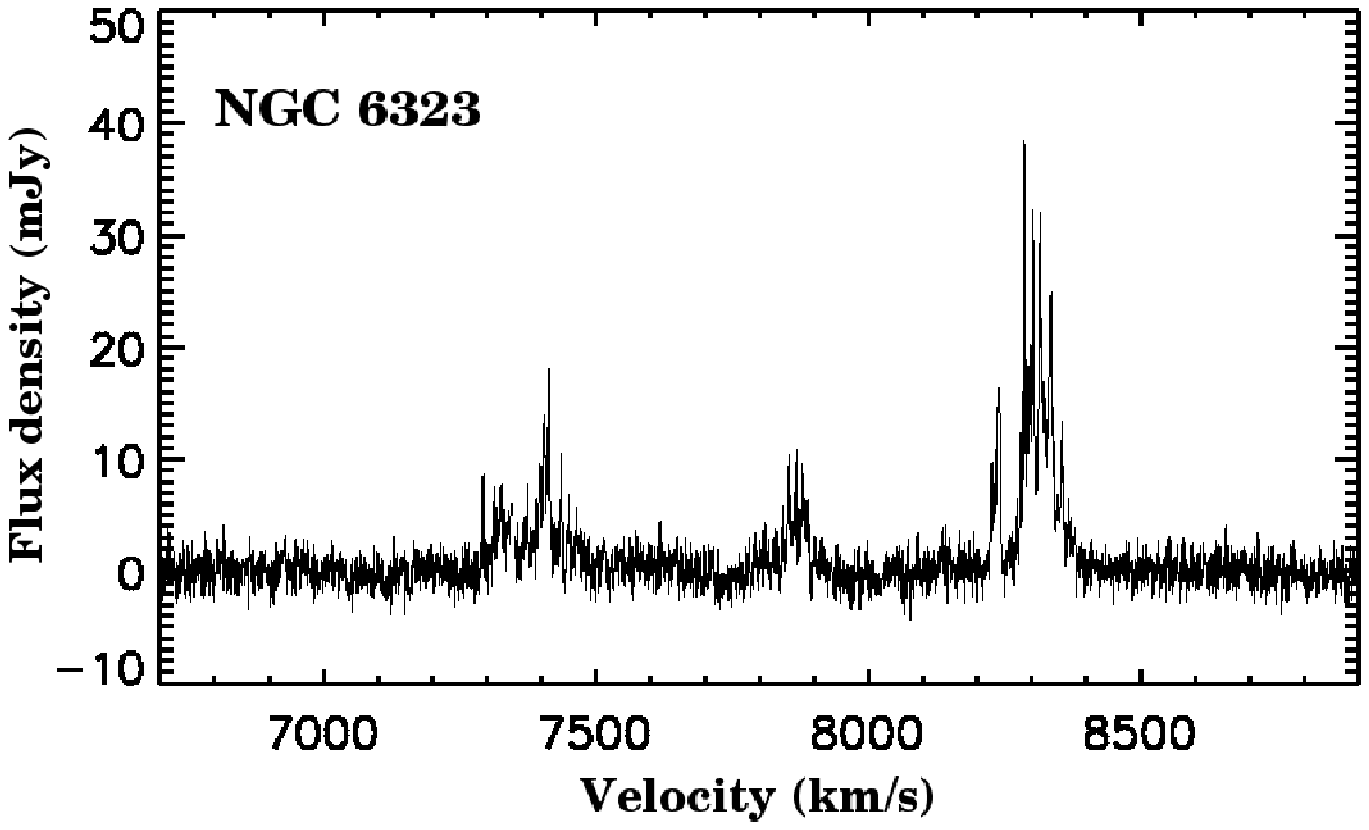} 
\vspace*{0.0 cm} 
\caption{A characteristic H$_{2}$O maser spectrum of NGC 6323 obtained on 2007 April 6. 
The vertical axis is flux density in mJy and the horizontal axis
is Local Standard of Rest velocity (optical definition). 
Note that the systemic maser emission between $\approx7800$ and $7900$ \kms\ is
weak ($<10$ mJy).}
\label{figure:spectrum}
\end{center} 
\end{figure}

We observed the H$_{2}$O maser in NGC 6323 with the GBT at 21 epochs
between 2006 October 30 and 2009 May 19. Except during the summer
months when the humidity makes observations at 22 GHz inefficient,
we took a spectrum on a monthly timescale. For these observations,
we followed the same observing settings and data reduction procedures
as in Braatz et al. (2010). Table \ref{table:dates} shows the observing date and
sensitivity for each observation. Figure \ref{figure:spectrum} shows a representative
H$_{2}$O maser spectrum for NGC 6323.                                   

\begin{deluxetable}{clcccc} 
\tabletypesize{\scriptsize} 
\tablewidth{0 pt} 
\tablecaption{GBT Observing dates and sensitivities for NGC 6323} 
\tablehead{ 
\colhead{Epoch} & \colhead{Date}      & \colhead{Day Number} 
 & \colhead{T$_{sys}$ (K)}  & 
\colhead{rms Noise (mJy)} & Period} 
\startdata 
0 & 2006 October 30  &   0  & 42.4 & 2.2 & A \\ 
1 & 2006 December 2  &  33  & 36.0 & 1.4  & A\\ 
2 & 2007 February 22 & 115  & 44.0 & 2.2  & A \\ 
3 & 2007 April 6     & 158  & 35.9 &1.8 & A\\ 
4 & 2007 October 29  & 364  & 34.3 &1.4 & B\\ 
5 & 2007 November 28 & 394  & 35.7 &1.6 & B\\ 
6 & 2007 December 26 & 422  & 55.0 &2.9 & B\\ 
7 & 2008 February 2  & 460  & 39.3 &1.5 & B \\ 
8 & 2008 February 29 & 487  & 44.0 &1.7 & B\\ 
9 & 2008 March 25    & 512  & 34.2 &1.3 & B\\ 
10 & 2008 April 24    & 542  & 56.5 &1.8 & B\\ 
11 & 2008 May 6       & 554  & 41.1 &2.4 & B \\ 
12 & 2008 May 29      & 577  & 49.4 &2.0 & B\\ 
13 & 2008 September 29 & 700  & 47.7 &2.2 & C\\ 
14 & 2008 October 31  &  732  & 43.5 &1.7 & C\\ 
15 & 2008 November 28 &  760  & 35.9 &1.1 & C\\ 
16 & 2008 December 29 &  791  & 35.2 &1.3 & C\\ 
17 & 2009 January 30  &  823  & 33.8 &1.1 & C\\ 
18 & 2009 March 4     &  856  & 31.5 &1.2 & C\\ 
19 & 2009 March 31    &  883  & 38.2 &1.3 & C\\ 
20 & 2009 May 19      &  932 & 35.3  & 1.5  & C \\ 
\enddata 
\label{table:dates}
\tablecomments{The sensitivities are calculated without performing
Hanning smoothing to the spectra and are based on 0.33 km~s$^{-1}$
channels. We label Period A, B, and C to those times when we have
continuous observations on a monthly timescale. These periods are
separated by summer months during which the humidity makes observations
inefficient. }                                                          
 
\end{deluxetable} 
 
\subsection{VLBI Data} 

We observed NGC 6323 with thirteen 12-hour tracks of VLBI observations
between 2007 and 2009 using the Very Long Baseline Array (VLBA)\footnote{The VLBA is a facility of the National Radio Astronomy Observatory, 
which is operated by the Associated Universities, Inc. under a cooperative
agreement with the National Science Foundation (NSF).}, 
augmented by the 100-m Green Bank Telescope and the 100-m Effelsberg telescope
\footnote{The Effelsberg 100-m telescope is a facility of the Max-Planck-Institut
f\"ur Radioastronomie.}. 
These observations have been presented in Kuo et al. (2011),
who used the inteferometric maps to estimate the supermassive black hole mass,
using a distance given by assuming $\Ho=73$ \kmsmpc\ and a recessional velocity
of 7848 \kms.

We have improved on the previous analysis by dividing the calibrated interferometer
(u,v)-data into yearly groups, periods A, B, and C (see Table \ref{table:dates}), merging data
within a group, and then imaging.  We could not combine all the data when
imaging, since most maser spots changed amplitude significantly over the 
three year span of the observations.  By comparing maser positions from different
periods, we can estimate the magnitude of systematic position offsets in the maser astrometry,
owing to independent phase and delay calibration. When comparing maser positions, we use the data from period C as the
reference because we have the most complete data in this period and this data dominate the sensitivity of the entire dataset. 
We found that there is no systematic offset between maser positions measured in period A and C, and the average maser positions from these two periods agree within 2 $\mu$as. However, when comparing positions measured from periods B and C, we found systematic position offsets for the systemic and blueshifted masers. There is no position offset for the redshifted masers in period B and C as expected because all maser positions are referenced to the average position of the redshifted masers as a result of the self-calibration process.

The average position of the systemic masers in period B has an offset of $1\pm5$ $\mu$as in the easterly ($x$) direction and $10\pm8$ $\mu$as in the northerly ($y$) direction from that of the systemic masers in period C. 
For the blueshifted masers, the average offsets in x and y between period B and C are $7\pm4$ $\mu$as and $22\pm8$ $\mu$as, respectively. Since the systemic and blueshifed masers have average frequency offsets of 31 MHz and 64 MHz, respectively, from the average frequency of the spectral channels of the redshifted masers used for self-calibration, the systematic position offsets in the $y$ direction appear to be frequency dependent in nature, implying a residual error of $\sim$0.4 nsec in the multiband delay calibration for our VLBI data taken in period B. In order to reduce the systematic error caused by these systematic position offsets in the model fitting described in Section 4, we corrected for these offsets in the maser positions measured in period B before performing disk modeling and Hubble constant determination.  


Next, we averaged the maser positions from the three periods to further increase
position precision.  Evidence from other (stronger) megamasers (NGC 4258 and UGC 3789)
supports the assumption that over time periods of a few years, the location and
acceleration of maser spots at any given velocity are nearly constant, even though their amplitudes can
vary.  The physical model of megamasers that is best supported by all 
observations (e.g. Humphreys et al. 2008, Reid et al. 2013, Kuo
et al. 2013) is that there is spiral structure in the accretion disk, and
typically one or two arms produce detectable maser emission.   Since, we observe 
the masers for only about 0.1\% of a maser orbital period, little
change in position or acceleration would be expected. 
In Table \ref{table:data}, we show the positions and velocities of all maser spots
used for the disk modeling.  (Throughout this paper, maser velocities
mentioned are referenced to the Local Standard of Rest (LSR), using
the optical velocity convention, even though full relativisitic corrections
are used when modeling).

\begin{deluxetable}{lrrrrrr} 
\tablewidth{0 pt} 
\tablecaption{Maser Position, Velocity, and Acceleration Data for
NGC 6323}                                                               
\tablehead{ 
\colhead{$V_{\rm op}$\tablenotemark{a}} & \colhead{$\Theta_{x}$\tablenotemark{b}}
& \colhead{$\sigma_{\Theta_{x}}$\tablenotemark{b}}  & \colhead{$\Theta_{y}$\tablenotemark{b}}
& \colhead{$\sigma_{\Theta_{y}}$\tablenotemark{b}}  & \colhead{$A$\tablenotemark{c}}
& \colhead{$\sigma_{A}$\tablenotemark{c}}                               
\\ 
\colhead{(km~s$^{-1}$)}  &\colhead{(mas)} 
&\colhead{(mas)}         &\colhead{(mas)} 
& \colhead{(mas)}  & \colhead{(km~s$^{-1}$~yr$^{-1}$)} & \colhead{(km~s$^{-1}$~yr$^{-1}$)}
}                                                                       
\startdata 
 8372.94  &  $-$0.021  &   0.009  &  $-$0.308  &   0.019  &     ...
&  ...  \\                                                              
 8365.82  &  $-$0.023  &   0.007  &  $-$0.274  &   0.014  &     ...
&   ...  \\                                                             
 8362.25  &  $-$0.035  &   0.010  &  $-$0.338  &   0.021  &     ...
&   ...  \\                                                             
 8358.69  &  $-$0.033  &   0.007  &  $-$0.323  &   0.012  &     ...
&   ...  \\                                                             
 8355.13  &  $-$0.030  &   0.004  &  $-$0.316  &   0.006  &     ...
&   ...  \\                                                             
 8351.57  &  $-$0.026  &   0.003  &  $-$0.317  &   0.006  &     ...
&   ...  \\                                                             
 8348.01  &  $-$0.023  &   0.004  &  $-$0.341  &   0.007  &     ...
&   ...  \\                                                             
 8344.44  &  $-$0.031  &   0.008  &  $-$0.332  &   0.016  &     ...
&   ...  \\                                                             
 8340.88  &  $-$0.027  &   0.005  &  $-$0.353  &   0.010  &     ...
&   ...  \\                                                             
 8337.32  &  $-$0.048  &   0.002  &  $-$0.364  &   0.004  &     ...
&   ...  \\                                                             
 8333.76  &  $-$0.047  &   0.002  &  $-$0.370  &   0.003  &  $-$0.22
&     0.25  \\                                                          
 8330.20  &  $-$0.051  &   0.002  &  $-$0.369  &   0.004  &     ...
&   ...  \\                                                             
 8326.63  &  $-$0.049  &   0.002  &  $-$0.382  &   0.004  &  $-$0.58
&     0.36  \\                                                          
 8323.07  &  $-$0.047  &   0.002  &  $-$0.370  &   0.004  &  $-$0.40
&     0.38  \\                                                          
 8319.51  &  $-$0.051  &   0.002  &  $-$0.364  &   0.003  &  $-$0.04
&     0.36  \\                                                          
 8315.95  &  $-$0.048  &   0.002  &  $-$0.368  &   0.003  &     0.11
&     0.36  \\                                                          
 8312.38  &  $-$0.056  &   0.002  &  $-$0.384  &   0.003  &     0.07
&     0.41  \\                                                          
 8308.82  &  $-$0.056  &   0.004  &  $-$0.383  &   0.007  &  $-$0.17
&     0.41  \\                                                          
 8305.26  &  $-$0.065  &   0.002  &  $-$0.407  &   0.004  &     ...
&   ...  \\                                                             
 8301.70  &  $-$0.064  &   0.002  &  $-$0.418  &   0.003  &  $-$0.06
&     0.30  \\                                                          
 8298.13  &  $-$0.065  &   0.002  &  $-$0.425  &   0.004  &     ...
&   ...  \\                                                             
 8294.57  &  $-$0.065  &   0.002  &  $-$0.430  &   0.003  &  $-$0.04
&     0.27  \\                                                          
 8291.01  &  $-$0.062  &   0.002  &  $-$0.430  &   0.004  &     0.00
&     1.00  \\                                                          
 8287.45  &  $-$0.063  &   0.002  &  $-$0.432  &   0.003  &     0.11
&     0.24  \\                                                          
 8283.89  &  $-$0.062  &   0.002  &  $-$0.433  &   0.003  &     0.00
&     1.00  \\                                                          
 8280.32  &  $-$0.062  &   0.002  &  $-$0.416  &   0.003  &  $-$0.16
&     0.24  \\                                                          
 8276.76  &  $-$0.071  &   0.003  &  $-$0.456  &   0.006  &     ...
&   ...  \\                                                             
 8273.20  &  $-$0.068  &   0.006  &  $-$0.446  &   0.011  &     ...
&   ...  \\                                                             
 8269.64  &  $-$0.073  &   0.009  &  $-$0.426  &   0.015  &     ...
&   ...  \\                                                             
 8258.95  &  $-$0.071  &   0.006  &  $-$0.494  &   0.010  &     ...
&   ...  \\                                                             
 8241.14  &  $-$0.089  &   0.003  &  $-$0.516  &   0.006  &     ...
&   ...  \\                                                             
 8237.58  &  $-$0.090  &   0.002  &  $-$0.521  &   0.004  &     ...
&   ...  \\                                                             
 8234.01  &  $-$0.096  &   0.002  &  $-$0.511  &   0.005  &     ...
&   ...  \\                                                             
 8230.45  &  $-$0.098  &   0.004  &  $-$0.513  &   0.007  &     ...
&   ...  \\                                                             
 8226.89  &  $-$0.096  &   0.003  &  $-$0.537  &   0.006  &  $-$0.40
&     0.39  \\                                                          
 8223.33  &  $-$0.101  &   0.004  &  $-$0.543  &   0.007  &     ...
&   ...  \\                                                             
 7885.90  &  $-$0.009  &   0.005  &  $-$0.014  &   0.009  &     1.49
& 	    0.41  \\                                                         
 7882.35  &  $-$0.000  &   0.004  &  $-$0.016  &   0.007  &     0.84
&      0.26  \\                                                         
 7878.80  &    0.001   &   0.004  &  $-$0.016  &   0.007  & 	0.63
&      0.25  \\                                                         
 7875.25  &  $-$0.002  &   0.005  &  $-$0.010  &   0.008  & 	0.89
&      0.20  \\                                                         
 7868.14  &    0.002   &   0.004  &     0.000  &   0.006  & 	0.79
&      0.14  \\                                                         
 7864.59  &    0.004   &   0.005  &  $-$0.017  &   0.009  & 	1.12
&      0.15  \\                                                         
 7853.93  &  $-$0.002  &   0.004  &     0.000  &   0.008  & 	1.27
&      0.54  \\                                                         
 7850.38  &    0.001   &   0.006  &     0.028  &   0.011  & 	0.80
&      0.24  \\                                                         
 7846.83  &    0.002   &   0.004  &     0.009  &   0.005  & 	1.07
&      0.11  \\                                                         
 7843.27  &  $-$0.004  &   0.010  &  $-$0.030  &   0.023  & 	1.07
&      0.11  \\                                                         
 7804.19  &    0.011   &   0.012  &     0.016  &   0.021  & 	1.47
&      0.27  \\                                                         
 7465.43  &     0.120  &   0.005  &     0.550  &   0.009  &     0.05
&     0.24  \\                                                          
 7447.73  &     0.118  &   0.004  &     0.504  &   0.008  &  $-$0.11
&     0.24  \\                                                          
 7437.10  &     0.093  &   0.004  &     0.483  &   0.007  &     ...
&   ... \\                                                              
 7433.56  &     0.093  &   0.003  &     0.461  &   0.006  &     ...
&   ... \\                                                              
 7415.85  &     0.097  &   0.003  &     0.387  &   0.005  &     ...
&   ... \\                                                              
 7412.31  &     0.091  &   0.002  &     0.394  &   0.003  &     0.39
&     0.42  \\                                                          
 7408.77  &     0.087  &   0.003  &     0.405  &   0.005  &     ...
&   ... \\                                                              
 7405.23  &     0.090  &   0.002  &     0.406  &   0.005  &     0.04
&     0.42  \\                                                          
 7401.69  &     0.093  &   0.004  &     0.415  &   0.006  &     ...
&   ...  \\                                                             
 7398.15  &     0.090  &   0.004  &     0.412  &   0.007  &     ...
&   ...  \\                                                             
 7394.60  &     0.083  &   0.005  &     0.424  &   0.010  &     ...
&   ...  \\                                                             
 7373.36  &     0.070  &   0.007  &     0.345  &   0.013  &     ...
&   ...  \\                                                             
 7369.81  &     0.066  &   0.009  &     0.341  &   0.020  &     ...
&   ...  \\                                                             
 7345.02  &     0.058  &   0.006  &     0.342  &   0.012  &     ...
&   ...  \\                                                             
 7337.94  &     0.054  &   0.006  &     0.334  &   0.010  &     ...
&   ...  \\                                                             
 7330.86  &     0.049  &   0.008  &     0.314  &   0.015  &     ...
&   ...  \\                                                             
 7327.32  &     0.047  &   0.006  &     0.300  &   0.010  &     ...
&   ...  \\                                                             
 7323.78  &     0.055  &   0.005  &     0.286  &   0.010  &     ...
&   ...  \\                                                             
 7320.23  &     0.052  &   0.006  &     0.286  &   0.011  &     ...
&   ...  \\                                                             
 7316.69  &     0.058  &   0.007  &     0.266  &   0.014  &     ...
&   ...  \\                                                             
 7313.15  &     0.047  &   0.008  &     0.283  &   0.017  &     ...
&   ...  \\                                                             
\enddata 
\label{table:data}
\tablecomments{ } 
\tablenotetext{a}{Velocity referenced to the LSR and using the optical
definition (no relativistic corrections).}                              
\tablenotetext{b}{East-west and north-south position offsets and 
uncertainties. Position uncertainties reflect fitted random errors 
only. } 
\tablenotetext{c}{Measured or estimated acceleration and its uncertainty
for each maser component.  }                                             
 
\end{deluxetable} 
  
\section{Acceleration Analysis} 


Following Kuo et al. (2013), we adopted two approaches to
measure accelerations of maser spots: initial estimates from the \emph{eye-tracking}
method for high velocity masers followed by two methods of \emph{global least-squares
fitting} (Humphreys et al. 2008; Braatz et al. 2010; Reid
et al. 2013; Kuo et al. 2013) for systemic masers.  In essence, the
eye-tracking approach yields estimates of acceleration by fitting a
straight line to spectral peak velocities, identified by eye from the spectra, 
as a function of time.  The global least-squares fitting method 
fits the amplitudes for a range of channels in all selected spectra simultaneously,
with a model consisting of multiple Gaussian-shaped lines that drift linearly in time. 
 
\subsection{High Velocity Masers} 

In Figure 2, we plot the radial velocities of the high-velocity maser
peaks as a function of time.  Note that while the blueshifted and
redshifted maser lines are distributed over a velocity range of $\sim$200
km~s$^{-1}$, not all maser features are represented in Figure 2.
This is because maser lines blend with their neighboring maser features
significantly. To measure the accelerations of the maser features
seen in Figure 2, we first identify the lines that are persistent
in time and then fit a straight line to the data to measure the accelerations
directly. The uncertainty of the measurements is estimated by scaling
the fitting error by the square root of reduced $\chi^{2}$.

\begin{figure}[!htb] 
\begin{center} 
\vspace*{0.0 cm} 
\hspace*{0.0 cm} 
\includegraphics[angle=0, scale=0.65]{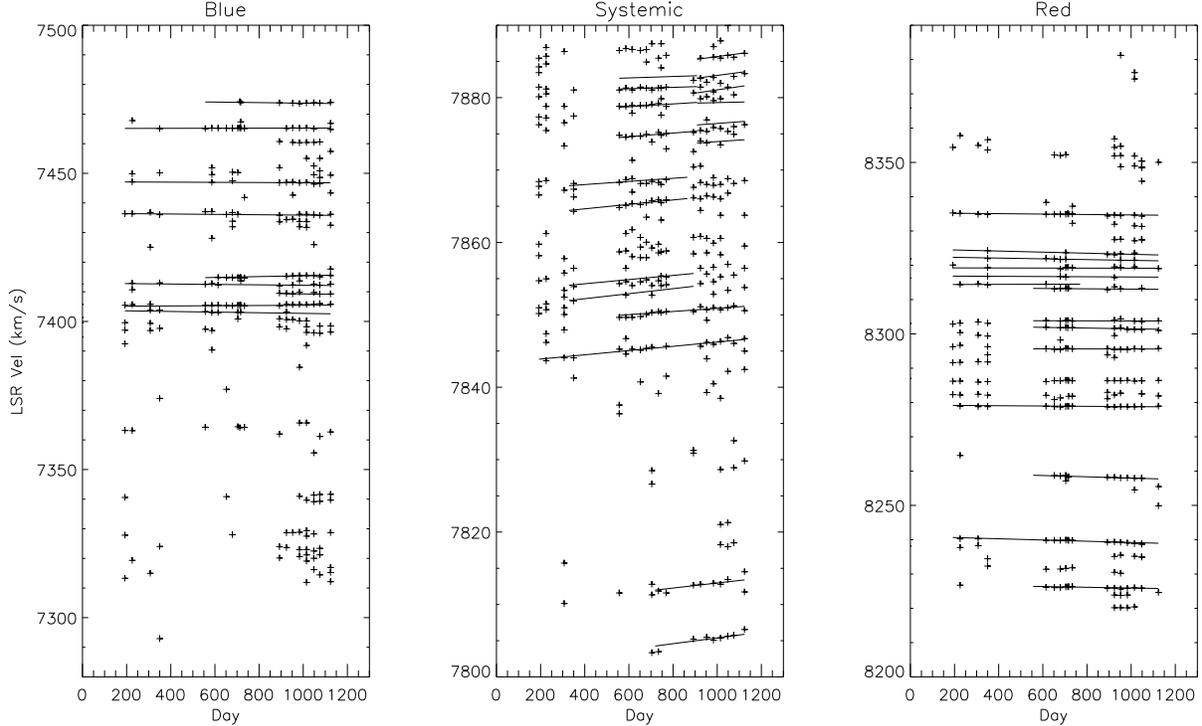} 
\vspace*{0.0 cm} 
\label{figure:byeye} 
\caption{We plot the radial velocities of NGC 6323 maser peaks as
a function of time (the crosses) for the blue-shifted masers (the
left panel), the systemic masers (the middle panel), and the red-shifted
masers (the right panel). On top of these plots, we overplot the
fitting results from the eye-tracking method for the high-velocity
masers and from the least-squares fitting for the systemic masers.
The data between Day 100 and 400 come from spectra taken in Period
A; the data between Day 500 to 800 from spectra in Period B; and
the data between Day 800 to 1200 are from spectra in Period C.  }       
\end{center}
\end{figure} 
 
The variance weighted average accelerations of the redshifted
and blueshifted masers are $-0.14$ and $-0.01$ \kmsyr\ respectively. 
The rms scatter of accelerations
of the redshifted and blueshifted masers are 0.26 and 0.21 \kmsyr.
The small accelerations and rms scatter indicate that the high velocity
masers are close to the mid-line of the accretion disk as expected.     
To account for possible uncertainty caused by line blending, 
we use the rms scatter of the acceleration measurement as an
estimate of the systematic error and include this error in the total
uncertainty before performing disk modeling described in Section 4.     
Fitted accelerations and uncertainties that include the systematic errors
are listed in Table \ref{table:data}.
 
\subsection{Systemic Masers} 
                        
\begin{figure}[!htb] 
\begin{center} 
\vspace*{0.0 cm} 
\hspace*{0.0 cm} 
\includegraphics[angle=0, scale=0.6]{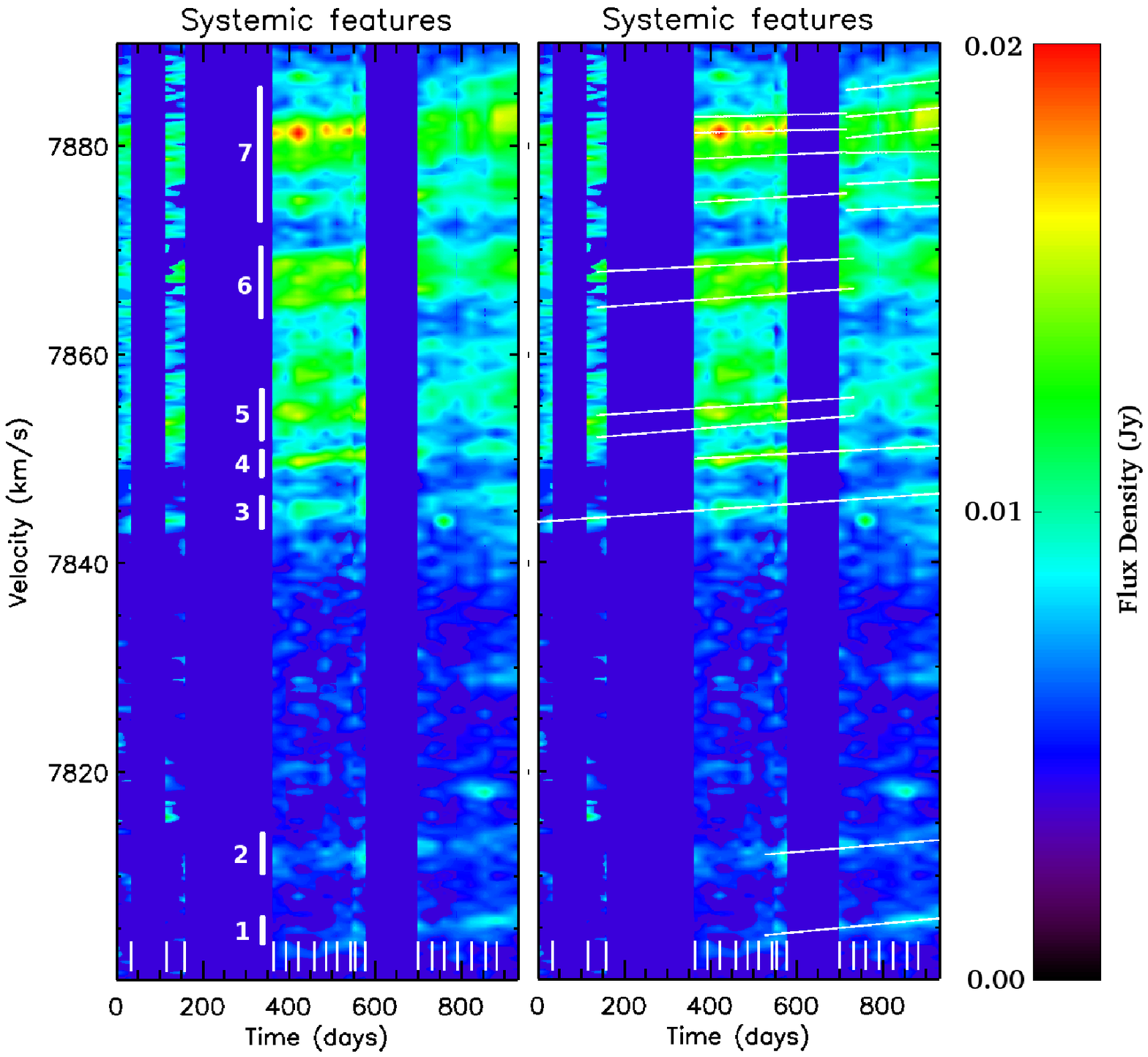} 
\vspace*{0.0 cm} 
\caption{\footnotesize Dynamic spectra of the systemic masers in NGC
6323.The left panel shows the maser flux density as a function of
time and velocity. The white ticks at the bottom of the figure show
the epochs at which we took the maser spectra. Except for epochs
separated by a summer gap, the maser flux at the time between these
epochs is obtained by linearly interpolating the flux densities measured
in consecutive epochs. No interpolation is used for consecutive spectra
that are separated by a summer gap (i.e. Day 160-360 and Day 580-700),
and these regions are intentionally left blank. The vertical white
lines in the figure show the velocity windows in which we can see
clear drifting pattern of maser lines and the accelerations of the
maser components within the window can be assumed to be approximately
the same. The numbers adjacent to the white lines are the group numbers
we assign to the velocity windows. In the right panel, we overplot
the straight lines that represent the best-fit accelerations from
the least-squares fitting on the plot shown in the left panel. } 
\label{figure:dynamic_spectra}       
\end{center} 
\end{figure}

The systemic features of NGC 6323 have low signal-to-noise ratios
(SNRs) and significant line-blending that make acceleration fitting
difficult.  To facilitate more reliable and stable acceleration fitting, 
we follow Kuo et al. (2013) and separate the maser lines into velocity sub-groups 
having similar apparent accelerations.  Assignment to an acceleration sub-group 
was done by examining \emph{dynamic spectra} to identify persistent patterns.
Dynamic spectra plot the flux density as color-coded intensity 
as a function of velocity versus time (see Figure \ref{figure:dynamic_spectra}).  
Flux densities were interpolated between monthly monitoring gaps, 
but not across longer (summer) gaps.  Based on the dynamic spectra, we identify
seven groups of masers that show coherent drifting patterns.  
Then, we estimated accelerations for the maser features in each group 
separately with global least-sqaures fitting.                         
 
When fitting the spectra for accelerations, we divide the data
into periods based on seeing coherent drifting patterns in the
dynamic spectra.   We show the epochs used in the
acceleration fitting for all maser groups in the 3rd column in Table
3. Note that we were unsuccessful fitting for masers in Group
5 and 6 between epochs 14 (Day 732) through 20 (Day 932), as a result 
of low SNR, severe line-blending, and especially the short time baseline.          
 
We followed \emph{Method 1} and \emph{Method 2} described in Reid et
al. (2013) to fit the accelerations of the systemic masers. 
These methods differ in the choice of initial parameter values and
the number of Gaussian components used.  Also, Method 1 allows each
Gaussian component to have independently estimated accelerations,
whereas Method 2 assumes maser accelerations are a linear function of 
velocity, defined as a single acceleration ($A_{sys}$) at the
center of the velocity range and its velocity derivative (d$A_{sys}$/dv). 
The difference between the measurements from these two methods
allows an estimate of uncertainty in measured acceleration, based
on fitting methodology.
 
\begin{figure}[!htb] 
\begin{center} 
\vspace*{0.0 cm} 
\hspace*{0.0 cm} 
\includegraphics[angle=0, scale=0.6]{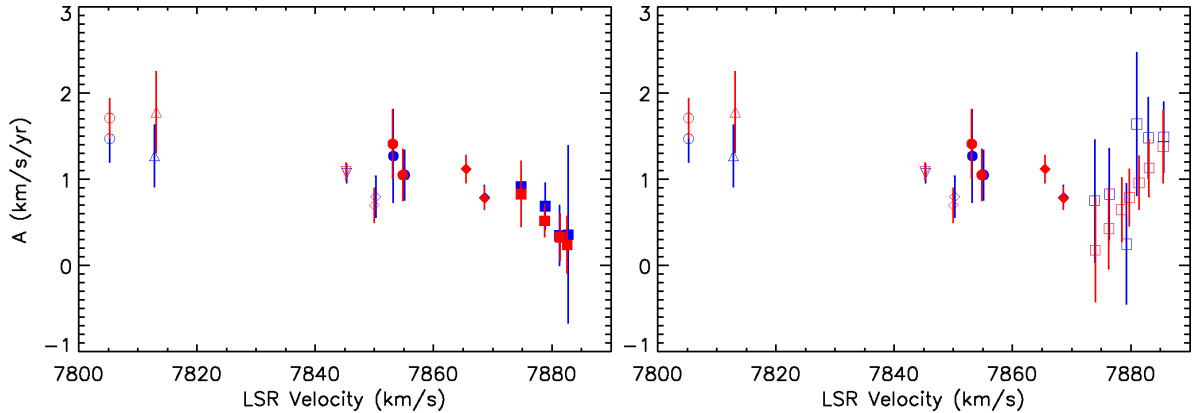} 
\vspace*{0.0 cm} 
\caption{\footnotesize
The left panel shows the centripetal acceleration of the
systemic masers measured from the global least-squares fitting method.
The red data points show the measurement assuming that the maser
acceleration is a linear function of maser velocity (Method 2, Sect.
3.2.1) whereas the blue ones show the measurement without such an
assumption (Method 1, Sect. 3.2.2). The measurements shown in the
right panel are the same as the ones shown in the left panel except
in the velocity range between 7873 km~s$^{-1}$ and 7888 km~s$^{-1}$.
We have different accelerations in this velocity window because we
measure the acceleration in two different sets of epochs. The left
panel show the measurements using spectra measured between epoch
0 and 13 (Day 0 $-$ Day 700; Group 7a; see also Table 4); the right
panel shows the measurement using data taken between epoch 14 and
20 (Day 732 $-$ Day 932; Group 7b). } 
\label{figure:accelerations}                                  
\end{center} 
\end{figure}

In Figure \ref{figure:accelerations}, we show the results of the 
acceleration measurements using both methods.  
Overall, the measurements from the two methods agree very well,
as do the accelerations obtained from the same group fitted
over different periods of time.    
For the disk fitting presented in Section 4, we adopted the
acceleraton measurements from  Method 1 (see Table \ref{table:accelerations}), 
because this method makes less restrictive assumptions.
Typical reduced $\chi^{2}_\nu$ values for these fits are 0.92-1.13 
for all maser groups except for Group 4, which typically had 
$\chi^{2}_\nu \approx 1.36$.  Only for the weak masers 
with velocites between 7882 and 7885 \kms\ do we see a marginally 
significant difference between measurements made between epochs
4--13 (in group 7a) and 14--20 (in group 7b).  For this velocity
range we adopt the variance weighted average of the two measurements.
In Section 4, we will explore the systematic uncertainty in \Ho caused by the slightly discrepant
acceleration measurements for masers in group 7a and 7b.

\begin{deluxetable}{ccrcccc} 
\tabletypesize{\scriptsize} 
\tablewidth{0 pt} 
\tablecaption{Acceleration Measurements for the Systemic Masers} 
\tablehead{ 
\colhead{Group}      & \colhead{Velocity} & \colhead{Epochs}  & \colhead{Ref.
Epoch}         & \colhead{Linewidth}  &                                 
\colhead{Acceleration}  & \colhead{$A_\sigma$}    \\ 
\colhead{} & \colhead{(km~s$^{-1}$)} & \colhead{}  & \colhead{} &
\colhead{Components} & \colhead{(km~s$^{-1}$~yr$^{-1}$)}    & \colhead{(km~s$^{-1}$~yr$^{-1}$)}
}                                                                       
\startdata 
1   &  7805.2  & 10$-$20   & 15 & 2.6     &  1.47    &  0.27    \\ 
2   &  7812.8  & 10$-$20   & 15 & 1.8     &  1.27    &  0.36    \\ 
3   &  7845.3  &  0$-$20   &  8 & 1.8     &  1.07    &  0.11    \\ 
4   &  7850.2  &  4$-$20   &  8 & 1.7     &  0.80    &  0.24    \\ 
5   &  7853.2  &  3$-$13   &  8 & 2.9     &  1.27    &  0.54    \\ 
5   &  7855.1  &  3$-$13   &  8 & 2.2     &  1.05    &  0.29    \\ 
6   &  7865.5  &  3$-$14   &  8 & 2.7     &  1.12    &  0.15    \\ 
6   &  7868.6  &  3$-$14   &  8 & 2.7     &  0.79    &  0.14    \\ 
7a   &  7874.8  &  4$-$13  &  8 & 2.9     &  0.92    &  0.23    \\ 
7a   &  7878.9  &  4$-$13  &  8 & 2.9     &  0.69    &  0.27    \\ 
7a   &  7881.3  &  4$-$13  &  8 & 1.8     &  0.35    &  0.35    \\ 
7a   &  7882.8  &  4$-$13  &  8 & 1.8     &  0.36    &  1.03    \\ 
7b   &  7873.9  &  4$-$13  & 16 & 2.7     &  0.75    &  0.71    \\ 
7b   &  7876.4  & 14$-$20  & 16 & 2.5     &  0.83    &  0.53    \\ 
7b   &  7879.3  & 14$-$20  & 16 & 2.4     &  0.25    &  0.70    \\ 
7b   &  7881.0  & 14$-$20  & 16 & 2.3     &  1.64    &  0.83    \\ 
7b   &  7883.0  & 14$-$20  & 16 & 2.2     &  1.48    &  0.47    \\ 
7b   &  7885.6  & 14$-$20  & 16 & 2.1     &  1.49    &  0.41    \\ 
 
\enddata 
\tablecomments{Col(1): maser group number; 
Col(2): the best-fit velocity of the model maser component, with a typical uncertainty of 0.25 \kms; 
Col(3): the epochs of the spectra used for fitting (for the dates see Table 1); 
Col(4): the reference epoch for velocity shown in Column(2); 
Col(5): the average linewidth; 
Col(6): the best-fit acceleration; and
Col(7): the uncertainty of the acceleration.
All fitted value come from accelertion fitting Method 1.}                                   
\label{table:accelerations}
\end{deluxetable}

\section{Modeling the Accretion Disk and Determining $H_{0}$} 
 
\subsection{Disk Modeling} 

The Hubble constant determination with the disk modeling method developed
in Reid et al. (2013) relies on modeling the sub-parsec scale maser
disk in three dimensions and adjusting model parameters to minimize 
the position, velocity, and acceleration residuals for each maser spot. 
Global model parameters include $H_{0}$, black hole mass ($M$), recession 
velocity of the galaxy ($V_{0}$), and other parameters that describe the
orientation and warping of the disk.
Key elements for estimating $H_{0}$ are the
Keplerian position-velocity (rotation) curve described by the high velocity 
masers, the centripetal accelerations of the systemic masers, and their angular
offsets and velocities relative to the dynamical center.   In essence,
the Keplerian rotation of the disk measured from high-velocity masers
determines $M/D$, where $D$ is the distance to the galaxy and $M$
is the black hole mass, whereas the position, velocity, and acceleration
information of the systemic masers give $M/D^{2}$ (see Kuo et al.
2013). Therefore, one can measure $D$ by solving the above two equations,
and $H_{0}$ can be directly inferred from $D$ and the galaxy's
Hubble flow speed, $V_{0}$, also obtained from the high-velocity rotation
curve.  We refer the readers to Reid et al. (2012) and Kuo
et al. (2013) for detailed information of the disk modeling.            
  
In the disk modeling, we associated maser positions and velocities
with accelerations measured from the GBT monitoring of maser 
spectra by choosing the VLBI channel with the velocity closest to that of an 
acceleration fit. For the high velocity masers, since there are fewer acceleration 
fits than VLBI measurements of maser position and velocity as results of the line-blending
effect and of using the eye-tracking method to measure the acceleration of high-velocity masers, 
not all VLBI measurements have corresponding accelerations.   For maser features without acceleration
measurements, we use only the position and velocity data in the disk modeling.
We show the input data (including the velocity, position, and acceleration
for each maser feature) for the disk modeling in Table \ref{table:data}.               
 
Rather than using formal fitting uncertainties for maser position
and velocity, which tend to be optimistic, we added estimates of
systematic uncertainty (``error floors'') in quadrature with the formal
uncertainties. For the $x-$ and $y-$data, we followed Reid et al. (2013) to adopt an error floor of 0.01 mas. 
For maser velocity, the precise value for the error floor is not important,
because it only affects the convergence rate of the disk fitting and does
not affect the best-fit result.  Therefore, as a rough estimate of the
systematic error for maser velocity, we adopt 1.8 \kms\ ($\approx$1/2 of the 
velocity channel width).                                       
 
We adopt a peculiar (deviation from a pure Hubble flow) velocity of 
$-285\pm163$ \kms\ for NGC 6323 (Masters; private communication).  This 
comes from galaxy flow models of the local supercluster (Masters 2005). 
Note that this peculiar velocity is relatively small, because toward
the Perseus-Pisces supercluster deviations from Hubble's flow are modest. 
Given the recession velocity of NGC 6323 is $\sim$7800 km~s$^{-1}$, the 
precise value of the peculiar velocity is unimportant as it contributes
only $\approx2$\% to the total $H_{0}$ uncertainty.
 
We modeled the maser disk in 3 dimensions with the same 10 global
parameters used in modeling the maser disk in UGC 3789 (Reid et al.
2013) and perform the model fitting with a Markov Chain Monte Carlo
(MCMC) approach (e.g. Geyer 1992; Gilks, Richardson \& Spiegelhalter
1996) to obtain the posteriori probability density distributions of the 
model parameters.  Optimum values of the model parameters were estimated
from the \emph{posteriori} probability density functions (PDFs),
marginalized over all other parameters.  To verify convergence of
the MCMC fitting, we adopted the power-spectrum method by Dunkley
et al. (2005). In this method, one measures the variance of the 
mean of the probability distribution and the correlation length of
the MCMC chain from its power spectrum, followed by evaluation of
the degree of convergence based on these two parameters. A MCMC chain
is considered to have converged if $r<0.01$ and $j^{*}\ge20$. Here, 
$r$ is the convergence ratio, defined as the ratio between the variance of the sample mean and variance
of the underlying probability distribution ($\sigma_{\bar{x}}^{2}$/$\sigma_{0}^{2}$), and $j^{*} = k^{*}(N/2\pi)$ indicates
the region in the power spectrum (i.e. the wavenumber $k^{*}$) of the MCMC chain where the chain
starts to deviate from the white noise regime (see Dunkley et al.
2005 for details). $N$ in the above expression for $j^{*}$ is the total number of samples in the MCMC chain. Finally, our estimates of the best global parameter values come 
from the marginalized PDFs; we adopted the median of the marginalized PDF, 
with the uncertainties showing the 16th and 84th percentiles 
(spanning the 68\% confidence interval).
  
\subsection{$H_{0}$ Estimation}

To fully explore the parameter space, we ran 10 independent MCMC
chains with different starting conditions. In particular, we choose
the starting value for $H_{0}$ uniformly in the range 60 $<$ $H_{0}$
$<$ 80 km~s$^{-1}$~Mpc$^{-1}$. The length of each independent run
was set to be 10$^{8}$ MCMC trials. The combined MCMC chains from
the 10 indepedent runs fully satisfied the convergence criteria set
by Dunkley et al. (2005), with convergence ratio $r=0.0007$
and $j^{*}=33$.                                                  
 
\begin{figure}[!htb] 
\begin{center} 
\vspace*{0.0 cm} 
\hspace*{0.0 cm} 
\includegraphics[angle=0, scale=0.7]{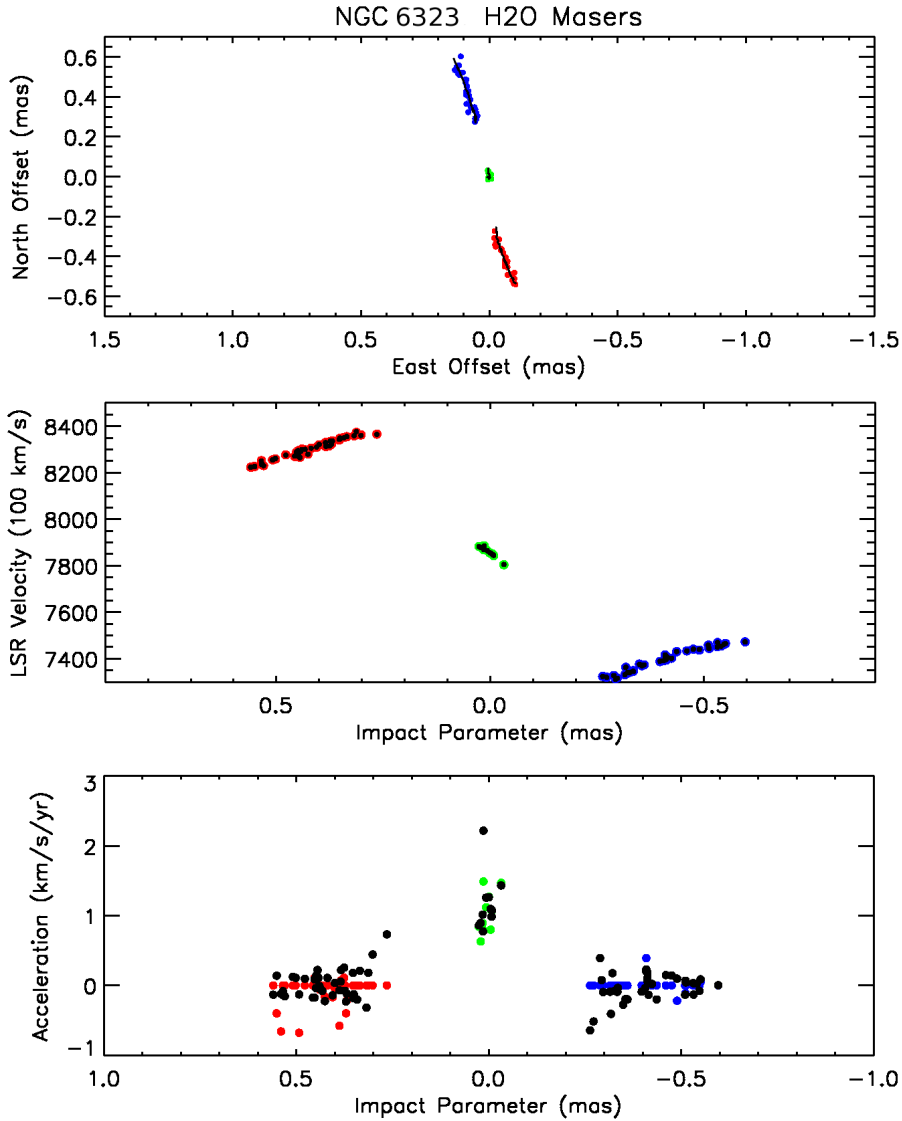} 
\vspace*{0.0 cm} 
\caption{\footnotesize
Data (colored dots) and best-fit model (lines and black
dots). Top panel: Positions on the sky.                                 
Middle panel: LSR velocity versus position along the disk. Bottom
panel: Accelerations versus impact parameter. In all three panels,      
the red, green, and blue dots show the redshifted masers, the systemic
masers, and the blueshifted masers, respectively. Note that in the bottom panel, there are
three redshifted maser spots with acceleration below $-$0.5 \kmsyr, which are larger than the typical acceleration (i.e. $\sim$0 \kmsyr)
for high-velocity masers and these values are very likely a result of severe line-blending. Excluding these
maser features does not change the \Ho from the disk modeling because the \Ho determination is dominated
by the systemic masers. The accelerations of the high-velocity masers
play a negligible role in the \Ho measurement.} 
\label{figure:data_model}                     
\end{center} 
\end{figure}

\begin{figure}[!htb] 
\begin{center} 
\vspace*{0.0 cm} 
\hspace*{-0.5 cm} 
\includegraphics[angle=0, scale=0.6]{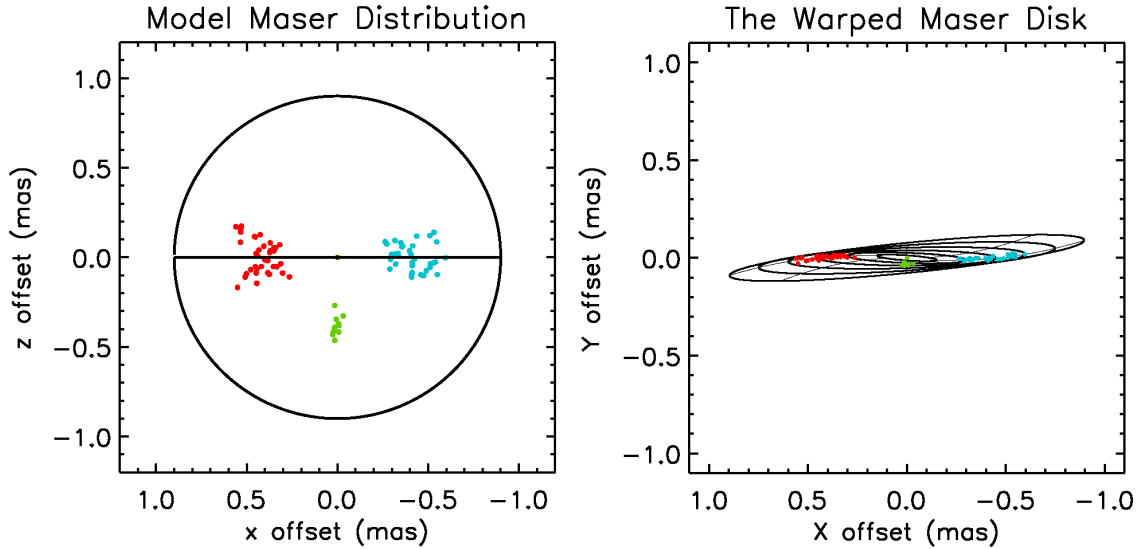} 
\vspace*{0.0 cm} 
\caption{\footnotesize 
The left panel shows the maser distribution of our best-fit
model for NGC 6323 from the overhead perspective. The right panel
shows the best-fit model from the observer's perspective with model
maser spots plotted on top of the warp model. The red, green, and
blue dots in both panels show the redshifted masers, the systemic
masers, and the blueshifted masers, respectively. In the right panel,
for the illustration purpose, we changed the observer's viewing angle
from 89$^{\circ}$ to 83$^{\circ}$ to show the degree of disk warping
more clearly.} 
\label{figure:overhead}                                                         
\end{center} 
\end{figure}

Figure \ref{figure:data_model} shows the results of the Bayesian fitting 
by comparing the most-probable MCMC trial model with the observed maser map, 
the position-velocity diagram, and the acceleration measurements.  
In Figure \ref{figure:overhead}, we show the $\emph{model}$ maser distribution 
from a view along the disk spin axis (left panel) and the warping structure with 
a nearly edge on view (right panel). In general, the model matches
the observations well, and the differences between the data and model
are consistent within realistic uncertainties except for a few high-velocity maser spots with relatively large 
acceleration (i.e. $\sim$$-$0.6 \kmsyr; see the bottom panel of Figure \ref{figure:data_model}). The total reduced $\chi^{2}$
($\chi^{2}_{\rm \nu}$) of the fit is 1.352 for 107 degrees of freedom
(291 data points).  We summarize the most-probable values for all model parameters 
in Table \ref{table:best_parameters}.  Note that since $\chi^{2}_{\rm
\nu}$ $>$ 1, we follow our conservative practice of inflating parameter 
uncertainties by $\sqrt{\chi^{2}_{\rm \nu}}$.  

\begin{figure}[!htb] 
\begin{center} 
\vspace*{0.0 cm} 
\hspace*{-0.5 cm} 
\includegraphics[angle=0, scale=0.6]{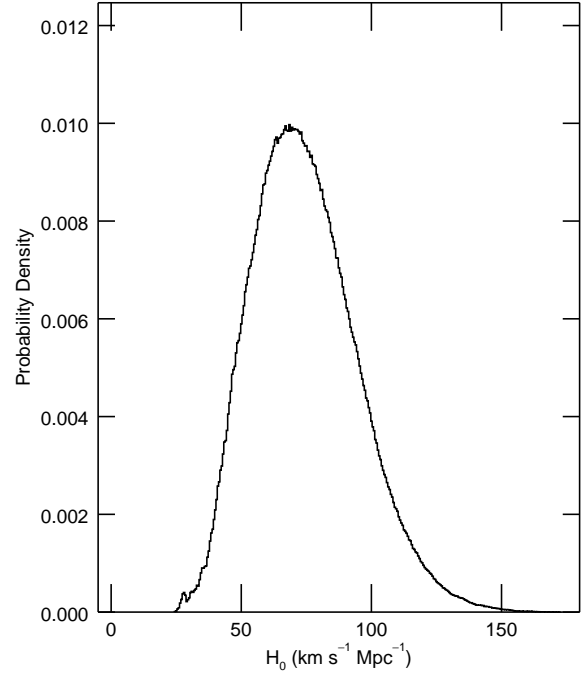} 
\vspace*{0.0 cm} 
\caption{Marginalized posteriori probability distribution of the
Hubble constant, $H_{0}$. The distribution for $H_{0}$ has a median
of 73 km~s$^{-1}$~Mpc$^{-1}$, and the 16th and 84th percentiles are
51 km~s$^{-1}$~Mpc$^{-1}$ and 99 km~s$^{-1}$~Mpc$^{-1}$, respectively.The
peak of the distribution is at $H_{0} = ~$67 km~s$^{-1}$~Mpc$^{-1}$.  } 
\label{figure:Ho}
\end{center} 
\end{figure}

We show the posteriori PDF for $H_{0}$ from the Bayesian fitting in 
Figure \ref{figure:Ho}. The PDF is slightly asymmetric about the peak of the 
distribution at $\Ho=67$ \kmsmpc.
The median of this PDF is 73 \kmsmpc\ and the 16th and 84th percentiles 
span 51 to 99 \kmsmpc.   From the fitted recessional velocity of $7853.4\pm2.2$ \kms,
the corresponding distance to NGC 6323 is $D=V_{0}/H_{0}$=107$^{+42}_{-29}$ Mpc. 
 
Note that in Section 3.2 we show that there is a marginally signifcant discrepancy
in the acceleration measurements for group 7a and group 7b, especially at the high end of the 
velocity window (see Figure \ref{figure:accelerations}). To explore the magnitude of the systematic error in \Ho caused by the
systematic uncertainty in acceleration, we perform the disk modeling again without including the systemic maser feature with velocity at 7885.9 km~s$^{-1}$ in Table \ref{table:data}. The resultant $H_{0}$ is 68$^{+26}_{-21}$ \kmsmpc. One can see that the $H_{0}$ decreases by 5 \kmsmpc and this represents the maximum systematic error in $H_{0}$ that the discrepant acceleration measurements for the maser features in group 7 at the high velocity end can introduce. Since this number is significantly smaller than the current measurement error in $H_{0}$, one can infer that the systematic error in the accelerations for group 7 has little impact on our current $H_{0}$ estimate which is dominated by measurement error.

\subsection{Sources of Uncertainties in $H_{0}$} 

In the context of a thin maser disk with well-measured Keplerian
rotation curve, as those cases shown in Kuo et al. (2011), the precision
of the $H_{0}$ measurement                                              
with the megamaser technique relies primarily upon the precision
of position and acceleration measurement of the systemic masers.
In addition, the spatial and velocity range over which systemic masers
can be detected also plays an important role.
 
For the maser disk in NGC 6323, the large $H_{0}$ uncertainty (compared
to other MCP galaxies) primarily results from the lower precision of the maser 
position measurements, owing to weaker maser emission and the fact that the 
maser disk orients in the north-south direction where we have the poorest angular 
resolution.  The accuracy of acceleration measurements is sufficient so that it 
is not a dominant contributor to the $H_{0}$ uncertainty.               
 
Besides position accuracy, the narrower velocity range of detectable
systemic masers in NGC 6323 is also a limiting factor for an accurate
$H_{0}$ measurement.  The velocity range covered by
the maser spots with position measurements to better than $\pm10$ $\mu$as 
is less than 50 km~s$^{-1}$, whereas in NGC 4258 (Humphreys et al. 2008) 
and UGC 3789 (Reid et al. 2013) the corresponding velocity ranges 
are $\approx100$ \kms.  Therefore, for  NGC 6323 our constraints on
$H_{0}$ are quite limited.

\newpage 
 

\begin{deluxetable}{cccccccccc} 
\tabletypesize{\scriptsize} 
\tablewidth{0 pt} 
\tablecaption{NGC 6323 H$_{2}$O Maser Model} 
\tablehead{ 
\colhead{Parameter}      & \colhead{Priors} & \colhead{Posterioris}
& \colhead{Units}                                                       
} 
\startdata 
$H_{0}$ & ...  & 73$^{+26}_{-22}$       & km~s$^{-1}$~Mpc$^{-1}$    \\ 
$V_{0}$ & ... & 7853.4$^{+2.1}_{-2.2}$   & km~s$^{-1}$               \\ 
$V_{p}$ & $-$285$^{+163}_{-163}$     & $-$282$^{+190}_{-188}$   
& km~s$^{-1}$               \\                                          
$M$     &  ...          & 0.94$^{+0.37}_{-0.26}$      & 10$^{7}$~$M_{\odot}$
\\                                                                      
x$_{0}$ &  ...         & 0.015$^{+0.002}_{-0.002}$    & mas     
\\                                                                      
y$_{0}$ &  ...         & 0.011$^{+0.002}_{-0.003}$    & mas     
\\                                                                      
$i_{0}$ &  ...          & 88.5$^{+0.7}_{-0.6}$         & deg    
\\                                                                      
$di/dr$ &  ...           & 6.0$^{+9.0}_{-6.2}$  
& deg~mas$^{-1}$             \\                                         
$p_{0}$ & ...      & 189.5$^{+0.2}_{-0.2}$       & deg          
\\                                                                      
$dp/dr$ & ...      & 13.2$^{+2.6}_{-2.6}$       &
deg~mas$^{-1}$              \\                                          
 
\enddata 
\tablecomments{ 
Parameters are as follows: Hubble constant ($H_{0}$), recession velocity of the galaxy
in optical convention relative to the Local Standard of Rest ($V_{sys}$), peculiar velocity with respect
to Hubble                                                               
flow in cosmic microwave background frame ($V_{p}$), black hole
mass ($M$), eastward                                                    
(x$_{0}$) and northward (y$_{0}$) position of black hole relative to a conveniently chosen point close
to the average position of all maser features, disk inclination
($i_{0}$) and                                                           
inclination warping (change of inclination with radius: $di/dr$), disk position                                                 
angle ($p_{0}$) and position angle warping (change of position position 
angle with radius: $dp/dr$). Flat priors were used,
except where                                                            
listed. Parameter values given in Table 4 were produced from binned
histograms for each parameter. The quoted values here are the medians
of the marginalized probability density functions, with the uncertainties showing the 16th
and 84th percentiles (i.e., the 68\% credible interval).                
} 
\label{table:best_parameters}
\end{deluxetable}

\section{Summary} 

This work presents the third \Ho\ estimate from the Megamaser Cosmology Project (MCP). 
Unfortunately, the low flux 
densities of the systemic masers in NGC 6323 preclude an accurate estimate
of \Ho. We discussed several approaches for handling low signal to noise
data, but conclude that with current telescope sensitivities the MCP should
concentrate on sources with maser spots stronger than $\approx20$ mJy.
However, if future radio telescopes can achieve significantly higher 
collecting areas and/or lower system temperatures, then we can make
substntially better measurements for galaxies such as NGC 6323.  We 
hope that the next generation Very Large Array will ultimately provide such an improvement.

\end{document}